\documentclass[12pt]{article}
\usepackage{psfrag}
\usepackage{amsmath}
\usepackage{amsthm}
\usepackage{amssymb}
\usepackage{epsfig}
\usepackage{euscript}
\usepackage{array}
\usepackage{bbold}
\usepackage{cancel}
\usepackage{mathtools}
\usepackage{empheq}
\usepackage{graphicx}
\usepackage{subfigure}
\usepackage{upgreek}
\usepackage{epstopdf}
\usepackage{booktabs}
\usepackage{mathrsfs}
\usepackage{amsbsy}

\usepackage{hyperref}
\usepackage{verbatim}

\theoremstyle{definition}

\theoremstyle{remark}

\newcounter{multieqs}

\newcommand{\lab}[1]{\label{#1}}
\newcommand{\prt}[1]{{\left( {#1} \right)}}

\newcommand{\be}{\begin{equation}}
\newcommand{\ee}{\end{equation}}
\newcommand{\eq}[1]{(\ref{#1})}
\newcommand{\bit}{\begin{itemize}}  \newcommand{\eit}{\end{itemize}}
\newcommand{\ben}{\begin{enumerate}}  \newcommand{\een}{\end{enumerate}}

\newcommand{\bm}[1]{\mbox{\boldmath $#1$}}
\newcommand{\rf}[1]{(\ref{#1})}

\def\bd{\begin{document}}
\def\ed{\end{document}}
\def\bea{\begin{eqnarray}}
\def\eea{\end{eqnarray}}
\let\bm=\bibitem

\def\la{\langle}
\def\ra{\rangle}

\def\npb#1#2#3{Nucl. Phys. {\bf{B#1}} #3 (#2)}
\def\plb#1#2#3{Phys. Lett. {\bf{#1B}} #3 (#2)}
\def\prl#1#2#3{Phys. Rev. Lett. {\bf{#1}} #3 (#2)}
\def\prd#1#2#3{Phys. Rev. {D bf{#1}} #3 (#2)}
\def\cmp#1#2#3{Comm. Math. Phys. {\bf{#1}} #3 (#2)}
\def\cqg#1#2#3{Class. Quantum Grav. {\bf{#1}} #3 (#2)}
\def\nppsa#1#2#3{Nucl. Phys. B (Proc. Suppl.) {\bf{#1A}}#3 (#2)}
\def\ap#1#2#3{Ann. of Phys. {\bf{#1}} #3 (#2)}
\def\ijmp#1#2#3{Int. J. Mod. Phys. {\bf{A#1}} #3 (#2)}
\def\rmp#1#2#3{Rev. Mod. Phys. {\bf{#1}} #3 (#2)}
\def\mpla#1#2#3{Mod. Phys. Lett. {\bf A#1} #3 (#2)}
\def\jhep#1#2#3{J. High Energy Phys. {\bf #1} #3 (#2)}
\def\atmp#1#2#3{Adv. Theor. Math. Phys. {\bf #1} #3 (#2)}

\def\N{{\cal N}}
\def\sst{\scriptscriptstyle}
\def\thetabar{\bar\theta}
\def\Tr{{\rm Tr}}
\def\one{\mbox{1 \kern-.59em {\rm l}}}

\def\a{\alpha}      \def\da{{\dot\alpha}}  \def\dA{{\dot A}}
\def\b{\beta}       \def\db{{\dot\beta}}
\def\g{\gamma}  \def\G{\Gamma}  \def\dc{{\dot\gamma}}
\def\d{\delta}  \def\D{\Delta}  \def\ddt{\dot\delta}
\def\e{\epsilon}
\def\ve{\varepsilon}
\def\uve{\upvarepsilon}
\def\f{\phi}    \def\F{\Phi}    \def\vvf{\f}
\def\vphi{\varphi}
\def\h{\eta}
\def\k{\kappa}
\def\l{\lambda} \def\L{\Lambda}
\def\m{\mu} \def\n{\nu}
\def\o{\omega}
\def\p{\pi} \def\P{\Pi}
\def\r{\rho}
\def\s{\sigma}  \def\S{\Sigma}
\def\t{\tau}
\def\th{\theta} \def\Th{\Theta} \def\vth{\vartheta}
\def\X{\Xeta}
\def\z{\zeta}

\def\na{\nabla}

\def\cA{{\mathscr A}} \def\cB{{\cal B}} \def\cC{{\cal C}}
\def\cD{{\cal D}} \def\cE{{\cal E}} \def\cF{{\cal F}}
\def\cG{{\cal G}} \def\cH{{\cal H}} \def\cI{{\cal I}}
\def\cJ{{\mathscr J}} \def\cK{{\cal K}} \def\cL{{\cal L}}
\def\cM{{\cal M}} \def\cN{{\cal N}} \def\cO{{\cal O}}
\def\cP{{\cal P}} \def\cQ{{\cal Q}} \def\cR{{\cal R}}
\def\cS{{\cal S}} \def\cT{{\cal T}} \def\cU{{\cal U}}
\def\cV{{\cal V}} \def\cW{{\cal W}} \def\cX{{\cal X}}
\def\cY{{\cal Y}} \def\cZ{{\cal Z}}

\def\ua{\underline{\alpha}}
\def\uc{\underline{\phantom{\alpha}}\!\!\!\gamma}
\def\um{\underline{\mu}}
\def\ud{\underline\delta}
\def\ue{\underline\epsilon}
\def\una{\underline a}\def\unA{\underline A}
\def\unb{\underline b}\def\unB{\underline B}
\def\unc{\underline c}\def\unC{\underline C}
\def\und{\underline d}\def\unD{\underline D}
\def\une{\underline e}\def\unE{\underline E}
\def\unf{\underline{\phantom{e}}\!\!\!\! f}\def\unF{\underline F}
\def\unm{\underline m}\def\unM{{\underline M}}
\def\unn{\underline n}\def\unN{{\underline N}}
\def\unp{\underline{\phantom{a}}\!\!\! p}\def\unP{\underline P}
\def\unq{\underline{\phantom{a}}\!\!\! q}
\def\unQ{\underline{\phantom{A}}\!\!\!\! Q}
\def\unH{\underline{H}}

\def\As {{A \hspace{-6.4pt} \slash}\;}
\def\bs {{b \hspace{-6.4pt} \slash}\;}
\def\Ds {{D \hspace{-6.4pt} \slash}\;}
\def\Gts {{\Gt \hspace{-6.4pt} \slash}\;}
\def\ds {{\del \hspace{-6.4pt} \slash}\;}
\def\ss {{\s \hspace{-6.4pt} \slash}\;}
\def\ks {{ k \hspace{-6.4pt} \slash}\;}
\def\ps {{p \hspace{-6.4pt} \slash}\;}
\def\xs {{x \hspace{-6.4pt} \slash}\;}
\def\pas {{{p_1} \hspace{-6.4pt} \slash}\;}
\def\pbs {{{p_2} \hspace{-6.4pt} \slash}\;}
\def\cFs {{{\cal F} \hspace{-6.4pt} \slash}\;}
\def\Dss {{D \hspace{-7.5pt} \slash}\;}
\def\dss {{\del \hspace{-7.0pt} \slash}\;}

\def\Ah{{\hat{A}}}
\def\Dh{{\hat{D}}}
\def\Gh{{\hat{G}}}
\def\Fh{{\hat{F}}}
\def\Ih{{\hat{I}}}
\def\Jh{{\hat{J}}}
\def\Kh{{\hat{K}}}
\def\Lh{{\hat{L}}}
\def\Ph{{\hat{P}}}
\def\Rh{{\hat{R}}}
\def\Vh{{\hat{V}}}
\def\Xh{{\hat{X}}}

\def\ah{{\hat{\a}}}
\def\bh{{\hat{\b}}}
\def\gh{{\hat{\g}}}
\def\dh{{\hat{\d}}}
\def\rh{{\hat{\r}}}
\def\hh{\hat{h}}
\def\uh{\hat{u}}
\def\xh{\hat{x}}
\def\yh{\hat{y}}
\def\ph{\hat{p}}
\def\xih{\hat{\xi}}
\def\chih{\hat{\chi}}
\def\Psih{\hat{\Psi}}
\def\phih{\hat{\phi}}

\def\psit{\tilde{\psi}}
\def\Psit{\tilde{\Psi}}
\def\Psibt{\tilde{\bar{Psi}}}

\def\st{\tilde{\sigma}}

\def\delt{\tilde{\delta}}
\def\Phit{\tilde{\Phi}}
\def\Phitb{\overline{\tilde{Phi}}}
\def\tht{\tilde{\th}}
\def\lt{\tilde{\l}}
\def\chit{\tilde{\chi}}
\def\phit{\tilde{\phi}}

\def\At{\tilde{A}}
\def\Bt{\tilde{B}}
\def\Ct{\tilde{C}}
\def\Dt{\tilde{D}}
\def\Et{\tilde{E}}
\def\Ft{\tilde{F}}
\def\Gt{\tilde{G}}
\def\Ht{\tilde{H}}
\def\It{\tilde{I}}
\def\Jt{\tilde{J}}
\def\Qt{\tilde{Q}}
\def\Rt{\tilde{R}}
\def\Mt{\tilde{M }}
\def\Nt{\tilde{N}}
\def\St{\tilde{S}}
\def\Vt{\tilde{V}}
\def\Xt{\tilde{X}}
\def\at{\tilde{a}}
\def\ct{\tilde{c}}
\def\dt{\tilde{d}}
\def\htt{\tilde{h}}
\def\ft{\tilde{f}}
\def\gt{\tilde{g}}
\def\pt{\tilde{p}}
\def\qt{\tilde{q}}
\def\vt{\tilde{v}}
\def\nt{\tilde{n}}
\def\ut{\tilde{u}}
\def\wt{\tilde{w}}
\def\zt{\tilde{z}}
\def\xt{\tilde{x}}
\def\yt{\tilde{y}}
\def\Psit{\tilde{\Psi}}
\def\vphit{\tilde{\varphi}}
\def\tD{\tilde{\D}}

\def\eb{\bar{\epsilon}}
\def\delb{\bar{\partial}}
\def\thb{\bar{\theta}}
\def\mub{\bar{\mu}}
\def\lamb{\bar{\l}}
\def\psib{\bar{\psi}}
\def\sb{\bar{\sigma}}
\def\xib{\bar{\xi}}
\def\chib{\bar{\chi}}

\def\Psib{\bar{\Psi}}
\def\Phib{\bar{\Phi}}
\def\Lamb{\bar{\Lambda}}
\def\Sb{{\overline \Sigma}}
\def\cb{\bar{c}}
\def\hb{\bar{h}}
\def\qb{\bar{q}}
\def\wb{\bar{w}}
\def\ub{\bar{u}}
\def\zb{{\bar{z}}}
\def\Hb{\bar{H}}
\def\Qb{{\bar Q}}
\def\Omegab{\overline{\Omega}}
\def\ob{\overline{\omega}}

\def\Ab{{\overline A}} \def\Bb{{\overline B}} \def\Cb{{\overline C}}
\def\Db{{\overline D}} \def\Eb{{\overline E}} \def\Fb{{\overline F}}
\def\Gb{{\overline G}}
\def\Ib{{\overline I}}
\def\Jb{{\overline J}} \def\Kb{{\overline K}} \def\Lb{{\overline L}}
\def\Mb{{\overline M}} \def\Nb{{\overline N}} \def\Ob{{\overline O}}
\def\Pb{{\overline P}}  \def\Rb{{\overline R}}
 \def\Tb{{\overline T}} \def\Ub{{\overline U}}
\def\Vb{{\overline V}} \def\Wb{{\overline W}} \def\Xb{{\overline X}}
\def\Yb{{\overline Y}} \def\Zb{{\overline Z}}

\def\fb{{\overline f}}
\def\gb{{\overline g}}
\def\mb{{\overline m}}
\def\lb{{\overline l}}
\def\yb{{\overline y}}

\def\ldel{{\overleftarrow{\del}}}
\def\rdel{{\overrightarrow{\del}}}
\def\ldeldel{{\overleftarrow{\del^2}}}
\def\rdeldel{{\overrightarrow{\del^2}}}
\def\ldelb{{\overleftarrow{\bar{\del}}}}
\def\rdelb{{\overrightarrow{\bar{\del}}}}

\def\ba{{\bf a}}
\def\bk{{\bf k}}
\def\bl{{\bf l}}
\def\bp{{\bf p}}
\def\bq{{\bf q}}
\def\br{{\bf r}}
\def\bt{{\bf t}}
\def\bu{{\bf u}}
\def\bv{{\bf v}}
\def\bx{{\bf x}}
\def\by{{\bf y}}
\def\bA{{\bf A}}
\def\bB{{\bf B}}
\def\bR{{\bf R}}
\def\bV{{\bf V}}

\def\bz{{\boldsymbol{\zeta}}}

\def\bone{{\bf 1}}

\def\va{{\vec a}}
\def\vk{{\vec k}}
\def\vp{{\vec p}}
\def\vq{{\vec q}}
\def\vx{{\vec x}}
\def\vy{{\vec y}}
\def\vu{{\vec u}}
\def\vv{{\vec v}}
\def \vH{{\vec H}}
\def \vg{{\vec g}}

\def\vs{{\vec \sigma}}
\def\vtau{{\vec \tau}}

\newcommand{\ov}[1]{\overrightarrow{#1}}

\def\frA{\mathfrak{A}}
\def\frB{\mathfrak{B}}
\def\frC{\mathfrak{C}}
\def\frD{\mathfrak{D}}
\def\frE{\mathfrak{E}}
\def\frF{\mathfrak{F}}
\def\frG{\mathfrak{G}}
\def\frH{\mathfrak{H}}
\def\frM{\mathfrak{M}}
\def\frN{\mathfrak{N}}
\def\frR{\mathfrak{R}}
\def\frW{\mathfrak{W}}

\def\fra{\mathfrak{a}}
\def\frb{\mathfrak{b}}
\def\frf{\mathfrak{f}}
\def\frg{\mathfrak{g}}
\def\frh{\mathfrak{h}}
\def\frl{\mathfrak{l}}
\def\frs{\mathfrak{s}}
\def\fri{\mathfrak{i}}
\def\frj{\mathfrak{j}}

\def\ma{\mathfrak{a}}
\def\mg{\mathfrak{g}}
\def\mh{\mathfrak{h}}
\def\mR{\mathfrak{R}}
\def\mN{\mathfrak{N}}

\newcommand{\nn}{{\nonumber}}

\def\d{\delta}\def\D{\Delta}\def\ddt{\dot\delta}

\def\pa{\partial} \def\del{\partial}
\def\xx{\times}
\def\uno{\mbox{1 \kern-.59em {\rm l}}}

\def\trp{^{\top}}
\def\inv{^{-1}}
\def\dag{\dagger}
\def\pr{^{\prime}}

\def\rar{\rightarrow}
\def\lar{\leftarrow}
\def\lrar{\leftrightarrow}

\newcommand{\0}{\,\!}      
\def\one{1\!\!1\,\,}
\def\im{\imath}
\def\jm{\jmath}

\newcommand{\tr}{\mbox{tr}}
\newcommand{\slsh}[1]{/ \!\!\!\! #1}

\def\vac{|0\rangle}
\def\lvac{\langle 0|}

\def\hlf{\frac{1}{2}}
\def\ove#1{\frac{1}{#1}}
\newcommand{\hot}[1]{\frac{#1}{2}}

\def\Box{\square}
\def\CC {\mathbb{C}}
\def\FF {\mathbb{F}}
\def\RR{\mathbb{R}}
\def\NN{\mathbb{N}}
\def\ZZ{\mathbb{Z}}
\def\bb#1{{\bf #1}}
\def\bcomment#1{}
\def\bfhat#1{{\bf \hat{#1}}}
\def\VEV#1{\left\langle #1\right\rangle}

\newcommand{\ex}[1]{{\rm e}^{#1}} \def\ii{{\rm i}}

\newcommand{\lrbrk}[1]{\left(#1\right)}
\newcommand{\lrsbrk}[1]{\left[#1\right]}
\newcommand{\sfrac}[2]{{\textstyle\frac{#1}{#2}}}

\def\stw{{\sqrt{2}}}

\def\rf {{\rm f}}
\def\ri {{\rm i}}
\def\rj {{\rm j}}
\def\rn {{\rm n}}
\def\rk {{\rm k}}
\def\rl {{\rm l}}
\def\rr {{\rm r}}
\def\rs {{\scriptscriptstyle \rm S}}
\def\rt {{\scriptscriptstyle \rm T}}

\def\rQ {{\scriptscriptstyle \rm \cQ}}
\def\rR {{\scriptscriptstyle \rm \cR}}

\def\cQb{{\cal \Qb}}
\def\cRb{{\cal \Rb}}
\def\cWb{{\cal \Wb}}

\def\fd {{\rm N}}
\def\afd {{\overline{\rm N}}}

\def \II {I\hspace{-.1em}I\hspace{.1em}}
\def \IIA {\mbox{\II A\hspace{.2em}}}
\def \IIB {\mbox{\II B\hspace{.2em}}}
\def \gs {g^s}
\def \ls {\lambda^s}

\def \I {{\cal I}}
\def \qs {q\hspace{-.53em}/\hspace{.15em}}
\def \ks {k\hspace{-.53em}/\hspace{.15em}}
\def \YM {{\mbox{\tiny YM}}}
\def \gym {g_{\YM}}

\def \Lc {\L_c}
\def\IR{\relax{\rm I\kern-.18em R}}
\def \id {{\bf 1}}

\def\cci{\ell}
\def\ccj{\ell'}

\def\bbq{\pmb{q}}

\begin{document}
\begin{titlepage}
\begin{flushright}
\hfill{ NCTS-TH/1601} 
\end{flushright}
\hfill

 \begin{center}

{\Large \bf AdS/dS CFT Correspondence
}\\[10mm]

{\bf Chong-Sun Chu${}^{1,2}$,  Dimitrios
Giataganas${}^1$}

{\itshape ${}^1$ Physics Division, National Center for Theoretical
  Sciences, \\
 National Tsing-Hua University, Hsinchu, 30013, Taiwan}\\[1mm]
{\itshape ${}^2$ Department of Physics, National Tsing-Hua
  University,  Hsinchu 30013, Taiwan}

{\small \sffamily
cschu@phys.nthu.edu.tw~, dgiataganas@phys.cts.nthu.edu.tw
\\
}
\end{center}

\date{\today}

\begin{abstract}

We propose and study  a  holographic dual of the type IIB
superstring theory of
AdS${}_5 \times S^5$ in terms of the $\cN=4$ superconformal Yang-Mills
theory on dS${}_4$. We use the bulk to boundary formalism to evaluate
the boundary
correlation functions
and verify
that it agrees with the expected result in dS conformal field
theory. 
The gauge theory is expected to be UV finite and
enjoy exact $SL(2,Z)$ strong-weak duality. As the string theory
Green-Schwarz sigma model carries an infinite number of classically
conserved charges, it also suggest that the superconformal
Yang-Mills theory is integrable and deserves further studies.

\end{abstract}

\end{titlepage}
\newpage


\section{Introduction}

de Sitter space plays a central role in modern theoretical physics. It
is not only relevant for the description of the late time
cosmology. It is also widely believed that the universe has
underwent a period of inflationary expansion described by a quasi-de
Sitter metric. As such it is of utmost  importance to understand the quantum
dynamics of de Sitter space. Despite intensive research,
e.g. \cite{witten-ds, strominger, KKLT} for some of the approaches, the
problem of quantum gravity in de Sitter space  remains open.
In fact even the much less ambitious problem of a quantum field theory dynamics
in de Sitter space is already quite a nontrivial problem. Although the
problem of the definition of Hamiltonian and particles for a
quantum field theory (QFT) in a generic curved spacetime
\cite{BD}
can be dealt with in perturbation theory for QFT in de Sitter space,
it is not necessary so in the strongly coupled regime, in which case
we also expect new phenomena may arise. Besides, the understanding of the
long time secular effects in de Sitter space is another important
problem that called for a treatment beyond the usual perturbation
scheme \cite{starobinsky,dRG,schwinger-dyson}.
It has been speculated that
\cite{poly,ford}
the infrared (IR) quantum effects in dS space could provide a screening
effect on the cosmological constant and offers a solution to the cosmological
constant problem.  One of the motivations of this work is the desire
for a better understanding of the dynamics of quantum field theories
in de Sitter spacetime beyond perturbation theory.

A powerful tool in this regard is the AdS/CFT correspondence
\cite{mal}.
The holographic nature of quantum gravity was originally
suggested by the discovery of the thermodynamic
nature of the black hole mechanics, most notably the Bekenstein-Hawking
formula for black hole entropy \cite{ber,hawking}.
The AdS/CFT correspondence
provides the first explicit example of how  holography
can be realized in string theory
\cite{mal,witten,GKP}.  Over the years, it has also been used as an
important tool to learn about the physical properties of various
quantum field theories in the strongly coupled
nonperturbative regime. See, for example, the reviews
\cite{r1,r2,r3}. In some cases,
when sufficient amount of supersymmetry or integrability is
present, even exact results are possible \cite{polchinski,r4}. In the case of
de Sitter space, gauge/gravity dual has been studied in
\cite{d1}-\cite{d24}.
In particular,
evidence of dynamical phase transition for confining gauge theory on
de Sitter space was found in the strongly coupled regime 
in \cite{d15};  entanglement of entropy for strongly coupled field theories
on de Sitter space with a gravity dual was computed 
in \cite{d17} and their results
suggested that the FRW cosmologies is contained in the field theory
description. One notes that for the kind of questions addressed
in these previous works, the gauge/gravity
correspondence was only needed to be considered in the generic sense without
having to spell out in details the involved
string theory and the boundary field theory.
Nevertheless it is certainly  interesting to have a concrete
duality so that one can ask precise questions of other kinds.
Another motivation of this work is to construct such a more precise
gauge/gravity correspondence in de Sitter spacetime.

The construction of global supersymmetric field theory in four dimensional
de Sitter spacetime is however impossible \cite{nogo1,nogo2}. First of all,
there does not exist Majorana Killing spinor on de Sitter spacetime,
which are necessary for the construction of supersymmetry. Moreover,
the usual de Sitter superalgebra has no unitary representation. These
no-go theorems are likely to be the reason why a holographic duality
involving a
supersymmetric field theory on de Sitter spacetime has not been
constructed. However it has been realized quite recently that the
no-go theorem can be bypassed if global superconformal symmetry is
considered instead of global supersymmetry.
In particular
the $\cN=4$ superconformal non-abelian Yang-Mills theory on dS${}_4$
has been constructed \cite{dzf}.  
The existence of this theory is also anticipated from the works of 
\cite{Hristov:2013spa,Cassani:2012ri}.
This superconformal Yang-Mills theory is
expected to enjoy exact $SU(2,2|4)$ supersymmetry \cite{dzf}. One
might also expect that
this superconformal theory may share some of the remarkable
properties such as integrability and S-duality
like its cousin, the $\cN=4$
supersymmetric Yang-Mills theory on Minkowski spacetime, which would
then allow
exact results to be obtained. Hence
it's gauge/gravity correspondence  would be a perfect laboratory to
see how some of the nontrivial results of the AdS/CFT correspondence would
extend in the presence of a time dependent background spacetime.
The desire to study the properties of the $\cN=4$ superconformal
Yang-Mills theory on de Sitter spacetime is another motivation of this work.

As it turns out, de Sitter spacetime can be obtained as the boundary
of the AdS spacetime. In the Poincare patch of the AdS space, a
boundary of Minkowski spacetime with a conformal structure was
created \cite{witten}. However one may also choose a dS-sliced
coordinates \eq{ds1} of the AdS${}_5$ spacetime and
obtain dS${}_4$ as the boundary manifold \cite{pope}.
Based on this 
observation
and the results of \cite{dzf}, we propose in this paper the AdS/dS CFT
correspondence:
Type IIB string theory on AdS${}_5 \times S^5$  with boundary
condition imposed on the boundary dS${}_4$ is dual
to the $\cN=4$ superconformal $SU(N)$ Yang-Mills theory on
dS${}_4$.

The plan of the paper is as follows.
In section 2 we review the global
definition of dS and AdS spacetime, and the
coordination of AdS which give rises to Minkowski or de Sitter
spacetime as boundaries. 
We also describe the $\cN=4$ superconformal Yang-Mills theory and some
general properties of conformal field theory on de Sitter spacetime.
In section 3, we introduce  the bulk to
boundary propagator formulation of the AdS/CFT correspondence \cite{witten,GKP},
generalized for a general bulk metric. In sections 4 and 5, we apply
this formulation to compute
the boundary correlators for 
conformal operators of scalar and  spin 1/2 types.
We show that the results obtained agree with the
de Sitter conformal field theory.
The paper is ended with further discussion in section 6.

\section{AdS/dS CFT Correspondence}

Our proposal is that
Type IIB string theory on AdS${}_5 \times S^5$  with boundary
condition specified on the boundary dS${}_4$ is dual
to the $\cN=4$ superconformal $SU(N)$ Yang-Mills theory on
dS${}_4$. Below let us spell out some of the basic elements of the
duality.

\subsection{dS Embedding in  AdS}


The $(d+1)$-dimensional Anti-de Sitter space AdS${}_{d+1}$ is a
maximally symmetric space with negative cosmological constant.
It can be most easily defined by an embedding
\be\label{ads1}
-X_0^2+X_1^2+\ldots+X_{d}^2-X_{d+1}^2=-L^2~,
\ee
in the $(d+2)$-dimensional flat space $\bR^{d+2}$ with the metric
$\eta_{\unM \unN} = \rm{diag} (-1, \mathbb{1}_d, -1)$. Here $L$ is the radius of
the AdS space and the cosmological constant is given by
\be\label{lc}
\L_0=-\frac{ d(d-1)}{2 L^2}~.
\ee
AdS${}_{d+1}$ is invariant under the group $SO(2,d)$ as both the
embedding metric and the embedding equations are invariant under this
transformation.
Similarly the de Sitter space dS is a maximally symmetric
space with positive cosmological constant.  For $d$-dimensional dS
space, it is given by the hyperboloid
\be
-Y_0^2+Y_1^2+\ldots+Y_{d}^2= L^2~,\lab{yconstr}
\ee
in the flat space $\bR^{d+1}$ with the metric
$\eta_{\unM \unN} = \rm{diag} (-1, \mathbb{1}_d)$. The cosmological constant
is
\be\label{lc1}
\L_1=\frac{ (d-1)(d-2)}{2 L^2}~,
\ee
and dS${}_d$ has the symmetry group $SO(1,d)$.



In the standard application of AdS/CFT correspondence,
one uses the Poincare
coordinates
\bea \label{poin}
&& X_0= \frac{r}{2}\left[ 1+ \frac{x^2- t^2 + L^2}{r^2}\right],
\qquad
 X_i = \frac{ L x_i}{r}, \quad i =1, \cdots d-1, \nn \\
&& X_d = \frac{r}{2}\left[ 1+ \frac{x^2- t^2 -L^2}{r^2}\right],
\qquad
X_{d+1} = \frac{L t}{r},
\eea
in which case the AdS metric takes the form
\be\label{adsp}
ds^2=\frac{L^2}{r^2}(d r^2 - dt^2 +d x_i^2), \quad r \geq 0.
\ee
It is clear that each constant $r$-slice describes
a copy of Minkowski space. The reason of the constraint
$r \geq 0$ is because of the singularity at $r =0$ and so the
metric \eq{adsp} can cover only half of the AdS space. 
Hence  the Poincare patch
describes a patch of the AdS
space with boundary consisting of a copy of the Minkowski space
$M_d$  at $r =0$,
together with a single point $P$ at $r = \infty$. This fact has
been of crucial
importance in the prescription of  \cite{witten,GKP}
for the realization of the holography of gravity in AdS
space.


It is interesting to note that the $(d+1)$-dimensional Anti-de Sitter
space  AdS${}_{d+1}$  also admit a
coordinate patch with $d$-dimensional de Sitter space
dS${}_d$ slicing. This fact has been used in the construction
of braneworld with cosmological constant \cite{pope}.
The embedding of dS${}_d$
can be realized with the following change of coordinates of
the AdS${}_{d+1}$:
\bea\lab{cc1}
X_{d+1} = L \cosh\frac{z}{L}~, \qquad\quad
X_\m= Y_\m\sinh\frac{z}{L}, \quad
\m=0,1,\ldots,d,
\eea
with $Y_\m$ satisfying \eq{yconstr} and describes  de Sitter space dS${}_d$.
In this coordinate patch, the AdS metric takes the form
\be\lab{ds1}
ds^2= dz^2+\sinh^2(\frac{z}{L}) \; ds^2_{dS}~, \quad z \geq 0.
\ee
The metric \eq{ds1} describes a portion of the AdS
space with boundary consisting of a copy of the de Sitter space
dS${}_d$ at $z =\infty$ , together with a single point at $z=0$.
An explicit solution of the constraint \eq{yconstr} is given by
\be\lab{cc2}
Y_0=\frac{\sinh H t}{H}-\frac{1}{2}H x_i^2 e^{-H t},\qquad
Y_i=x_i e^{-H t}, \qquad
Y_{d}=\frac{\cosh H t}{H}-\frac{1}{2}H x_i^2 e^{-H t},
\ee
with $H= 1/L$ and
$i = 1, \cdots, d-1$. This gives the dS${}_d$ metric in terms of the planar
coordinates $(t,x^i)$:
\be\lab{ds2}
ds^2_{dS} =-dt^2+e^{-2 H t} d x_i^2~ .
\ee

The normalized distance $P$
between two points $X$ and $X'$ in the ambient space
\be\label{inv1}
P(X,X'):=\frac{\eta_{\unM \unN} X^\unM X'^\unN}{L^2}~,
\ee
is a convenient quantity that can be used to express
the geodesic distance $\cD$ between any two points in the AdS or dS.
For time like separated points in AdS, $P$  is related to
the geodesic distance $\cD$ between two points $X, X'$ by
$ P =  - \cos (\cD/L)$.
For points in the same causal
diamond in dS,  the relation is  $P = \cos
(\cD/L)$.
For  AdS${}_{d+1}$, $P$ is
given by
\be
P_{\rm AdS}(X,X')=-\frac{1}{\xi}~,
\ee
where
\be\lab{geod-poin}
\xi^{-1} =\frac{r^2+r'^2+(x_i-x_i')^2-(t-t')^2}{2 r r'}
\ee
in the Poincare coordinates \eq{poin};
and
\be\label{geod-planar}
\xi^{-1} =  \cosh H z \,  \cosh H z' - \sinh Hz \, \sinh Hz'
 \times P_{dS} (x^\mu, x'{}^\mu)
\ee
in the dS planar coordinates \eq{cc1}, \eq{cc2}.
Here
\be
P_{\rm dS} := \cosh H(t-t') - \frac{e^{-H(t+t')}}{2} \; H^2
  (x_i-x_i')^2
\ee
is the normalized distance in dS${}_d$. Note that $P_{\rm dS}=1$ for coincident
points, therefore it is more convenient to consider the following
quantity
\be \label{sigma}
\s^2(x,x'):= e^{-H(t+t')} \prt{x_i-x_i'}^2
-\frac{\cosh H \prt{t-t'}-1}{H^2/2} ~
\ee
as a measurement of distance between any two points $(t,x_i), (t', x'_i)$ in
dS${}_d$.
In general it is
\be
\frac{P_{\rm dS} -1}{H^2} = -\frac{1}{2}\s^2.
\ee
The quantity $\s^2$ has the property that it coincides with the proper
distance in Minkowski spacetime in the flat space limit $H \to 0$,
\be
\s^2 \to |x-x'|^{2} := -(t-t')^2 + (x-x')^2 .
\ee
In terms of the conformal time  $x^0
= H^{-1}\exp (Ht)$, the dS metric \eq{ds2} can be written as
\be \label{ds3}
ds^2 = \frac{1}{H^2 x_0^2} \prt{-dx_0^2+ dx_i^2 }
\ee
and it is
\be \label{sigma1}
\s^2 = \frac{(x_\m -x'_\m)^2}{H^2 x_0 x'_0}~,
\ee
where indices are raised and lowered by the Minkowski metric $\eta_{\m\n}$.

\subsection{ $\cN=4$ superconformal Yang-Mills theory}

Let us review the $\cN=4$ superconformal Yang-Mills theory on dS${}_4$
constructed in \cite{dzf}.   There the metric is taken to be of the
form \eq{ds3}. Crucial to the construction of \cite{dzf} is the
existence of conformal Killing spinor on dS${}_4$.
Unlike the Killing spinor equation,
the conformal Killing spinor defined by the equation
\be
\prt{ D_\mu - \frac{1}{4} \g_\m \Dss} \e = 0
\ee
is compatible with the Majorana condition on spinor. This can be
solved and one obtain the conformal Killing spinors on dS${}_4$
\be
\e(x) = \frac{1}{\sqrt{H x^0}}(\eta_0 + x^\m \g_\m \eta_1 )~,
\ee
where $\eta_0, \eta_1$ are arbitrary Majorana spinors.
This gives $\cN=1$ superconformal symmetry in dS${}_4$ and
corresponds to a basis of 8 real supercharges.

The $\cN=4$ maximal superconformal Yang-Mills theory on dS${}_4$ contains the
gauge potentials $A_\m^a$, four Majorana gauginos $\l^a_\a$ and six
real scalars $X^{a}_i$, where the indices $a$ is in the adjoint of the gauge group
$SU(N)$. The Lagrangian is $\cL = \cL_2+ \cL_3 + \cL_4$ where
\bea
\cL_2 & = & - \Big[ 
\frac{1}{4} F_{\mu\nu}^a F^{\m\n a} + \lamb^{a\a} \g^\m D_\m P_L
\l^a_\a+
\frac{1}{2} D_\mu X_i^a D^\mu X^a_i + H^2 X^a_i X^a_i
\Big]~, \\
\cL_3 &=& - \frac{1}{2} f^{abc} X_i^a \Big[
C_i^{\a\b} \lamb^b_\a  P_L\l^c_\b +C_{i\a\b} \lamb^{b\a}P_R \l^{c\b}
\Big]~, \\
\cL_4 &=& -\frac{1}{4}f^{abc} f^{a'b'c'} X^b_i X^c_j X^{b'}_i X^{c'}_j~.
\eea
Here $P_L, P_R$ are chiral projectors, $C_i$ are the six `t
Hooft instanton matrices:
\bea
C_1 &=& \begin{pmatrix} 0 & \sigma_1\\ -\s_1 & 0 \end{pmatrix}~, \quad
C_2 = \begin{pmatrix} 0 & -\s_3\\ \s_3 & 0 \end{pmatrix}~, \quad
C_3 = \begin{pmatrix} i \s_2 & 0\\ 0& i\s_2 \end{pmatrix}~,\\
C_4 &=& -i  \begin{pmatrix} 0 & i\s_2\\ i\s_2 & 0 \end{pmatrix}~, \quad
C_5 = -i  \begin{pmatrix} 0 & 1\\ -1 & 0 \end{pmatrix}~, \quad
C_6 = -i  \begin{pmatrix} -i \s_2 & 0\\ 0& i\s_2 \end{pmatrix}
\eea
and $\s_i$ are the Pauli matrices. Note that $C_1, C_2, C_3$ are
real, $C_4, C_5, C_6$ are imaginary.

The action admits an $SU(4)$ R-symmetry
and the superconformal symmetry:
\bea
\d A_\m^a &=& - \eb^\a  \g_\mu P_L \l^a_\a - \eb_\a \g_\m P_R \l^{a\a}~, \\
\d X^a_i &=& -\eb_\a P_L C_i^{\a\b} \l_\b - \e^\a P_R C_{i\a\b}
\l^{a\b}~, \\
\d \l^a_\a &=& \frac{1}{2} \g^{\mu\nu} F_{\m\n}^a \e_\a -\g^\m D_\m X_i^a
(P_L C_i^{\a\b} \e_\b +P_R C_{i \a\b} \e^\b)
-\frac{1}{2} X_i^a (P_R C_i^{\a\b} \Dss \e_\b +P_L C_{i \a\b} \Dss
\e_\b) \nn\\
&& - \frac{1}{2}f^{abc} X_i^b X_j^c [(C_i C_j)^\a{}_\b P_R \e^\b
+ (C_i C_j)_\a{}^\b P_L \e_\b]~,
\eea
where $P_L\e_\a, P_R \e^\a$ are an $SU(4)$ quartet of Majorana
conformal Killing spinors. Here $(C_i C_j)^\a{}_\b := C_i^{\a\g} C_{j
  \g\b}$ and $ (C_i C_j)_\a{}^\b := C_{i\a\g} C^{\g\b}_j$.
Due to its large amount of superconformal symmetry, the theory is
expected to be UV finite. Since there is no massless minimal coupled
field, it is also expected that there is no IR divergence.
Hence the theory is expected to enjoy exact $SU(2,2|4)$ supersymmetry.
Adding a $\th$-term and restore the Yang-Mills coupling $g$, one
expect the theory also enjoy exact $SL(2,Z)$ strong-weak duality, just
as the type IIB superstring theory does.

The moduli space of the theory can be easily worked out. With fermions
and the gauge fields set to zero. The
equation of motion for the scalar fields read
\be
-D_\m D^\m X_i + 2 H^2 X_i - [X_j,[X_i,X_j]] =0~.
\ee
One class of solution (static) is given by product of fuzzy spaces described by
$SU(2)$. A more interesting  solution is
\be
X_i = e^{Ht} Z_i~,
\ee
where $Z_i$ are diagonal. This describes an expanding $R^6$
and is
expected from the form \eq{ds2} of the de Sitter metric used.

\subsection{Conformal field theory in dS space}

The  $SO(1,d)$ isometries of
de Sitter space dS${}_d$ is generated by the generators:
\be
L_{AB} = Y^A \frac{\del}{\del Y^B} - Y^B \frac{\del}{\del Y^A}, \quad
A, B = 0, 1, \cdots, d,
\ee
which acts linearly on the de Sitter hyperboloid.
In terms of the dS space conformal coordinates \eq{ds3},
the de Sitter isometries are generated by the
spatial rotations $J_{ij}$,  dilation $D$,
spatial translations $P_i$ and  special conformal
transformation $K_i$:
\bea
J_{ij} &=& -i L_{ij}= -i( x_i \del_j - x_j \del_i) ,
\quad  \mbox{where $i =1,\cdots, d-1$},
\\
D  &=& -i L_{0d} = -i x^\m \del_\m, \\
P_i &=& -i (L_{id} +L_{0 i}) = -i H^{-1}\del_i , \\
K_i &=& i (L_{id} - L_{0 i}) = -2iHx_i \, x^\m \del_\m  - iH
x^2 \del_i .
\eea
The corresponding finite transformations are:
\bea
x_i'& =& \L_i^j x_j, \qquad  \mbox{$\L \in  SO(d-1)$ rotations}, \\
x'{}^\m &=& \l x^\m,\\
x_i' &=& x_i +a_i, \\
x'{}^\m &=& \frac{x^\m + b^\m x^2}{1+2b_\m x^\m + b^2 x^2}, \quad b^\m
= (0, b^i).
\eea

As the de Sitter metric is related to the flat space Minkowski
metric by a Weyl factor, the conformal symmetry of dS${}_d$ is the same
$SO(2,d)$ as that of the Minkowski spacetime, and is obtained by
adding to the dS isometries the generators: 
the Lorentz boosts
$J_{0i}$, time translation $P^0$ and the special conformal transformation $K^0$. 
The corresponding finite transformations are:
\bea
x'{}^\m& =& \L^\m_\n x^\n, \qquad  \mbox{nonvanishing $\L^0_i, \L^i_0$
  = Lorentz boost}, \\
x'{}^0 &=& x^0 +a^0, \\
x'{}^\m &=& \frac{x^\m + b^\m x^2}{1+2b_\m x^\m + b^2 x^2}, \quad b^\m
= (b^0, 0)
\eea
and the metric transforms as
\be
ds^2 \to ds'{}^2 = \L^2(x) ds^2, 
\ee
where, respectively,
\be
\L(x)^2 = \prt{\frac{x_0}{\L^0_\m x^\m}}^2, \qquad 
\prt{\frac{x_0}{x_0+a_0}}^2, \qquad 1 . 
\ee
As usual, the special
conformal transformations can be constructed out of translations and
the inversion
\be \label{inv}
x^\m \to x'{}^\m = \frac{x^\m}{x^2} := \prt{\frac{1}{x} }^\m.
\ee
Inversion is an isometry and induces an $SO(1,d-1)$ rotation on  vector
\be
\frac{\del}{\del x'{}^\m} = x^2 M_\m^\n(x) \frac{\del}{\del x^\n} ,
\ee
where
\be
M_\m^\n(x) := \d_\m{}^\n - 2\frac{x_\m x^\n}{x^2}
\ee
and satisfies $M_\m^\a M_\n^\b \eta_{\a\b} =
\eta_{\m\n}$. For a spinor in fundamental representation, inversion induces the
transformation
\be
\psi'(x) =
S_\a^\b(x) \psi_\b\prt{\frac{1}{x}},
\ee
where the  matrix $ S^\a_\b(x)$ satisfies
\be
S^\dag(x) \G^\m S(x) = M^\m_\n \G^\n
\ee
and is given by
\be \lab{smatrix1}
S(x) = \frac{\G_\m x^\m}{|x|}.
\ee

In a conformal field theory,
scalar (with respect to the rotation group) 
operator $\cO$ of conformal dimension $\D$ satisfies
under conformal transformation $x\to x'$ as
\be
\quad\cO'(x') = \frac{1}{|\L(x)|^{\D}} \cO(x).
\ee
For two such operators, 
dS invariance implies that their
$2$ point function must be function of the geodesic distance
$\s(x,y)^2$. Furthermore, invariance under the conformal
transformations implies  that
\be
\langle \cO_1 (x) \cO_2(y) \rangle = \left\{
\begin{array}{ll} 
\dfrac{C_{12}}{\s(x,y)^{2\D}} & \mbox{if $\D_1 = \D_2 = \D$}, \\
0  &   \mbox{if $\D_1 \neq \D_2$}.
\end{array}\right. 
\ee
As for operator 
$\cO^\a(x)$
of dimension $\D$ in spin 1/2 representation, it 
satisfies under translation in $x^0$:
\be
\cO'{}^\a(x') = \frac{1}{|\L(x)|^{\D}} S^\a_\b(x) \cO^\b(x^\m).
\ee
Together with it's property under inversion
\be
\cO'{}^\a(x') = S^\a_\b(x) \cO^\b(x^\m),
\ee
one can easily  show that the two point function 
for any two such operators is 
\be \label{OOD}
\langle \cO^\dag_{1\,\a}(x) \cO_2^\b(y) \rangle =
\frac{D_\a{}^\b(x,y)}{\s(x,y)^{2\D}} \d_{\D_1, \D_2}, 
\ee
where $\D= \D_1$. Here $D$  satisfies the relation
\be
D(x,y) = S^\dag(x) D\prt{\frac{1}{x},\frac{1}{y}} S(y)
\ee
and is given by
\be
D_\a{}^\b(x,y) = \frac{(x-y)_\m \G^\m}{|x-y|}. 
\ee
Constraints on 3 and higher point functions of dS CFT
can be similarly worked out.

\section{Bulk Propagators and
Boundary Correlators: General Setup} \label{b2b-general}

One of the very basic tools for the study of gauge/gravity
correspondence is the bulk to boundary formalism
\cite{witten, GKP}. In this section
we develop this formalism for a general bulk metric. Most of what we do in
section \ref{b2b-scalar} is a simple generalization of
known results in the literature. The general
result obtained in section \ref{b2b-fermion} is new.

Consider a $(d+1)$-dimensional manifold $\cM$ with boundary and the metric
\be
ds^2=g_{MN} dy^M dy^N~.
\ee
Without loss of generality, we assume that
the metric has an expansion of the form near the boundary,
\be \label{bdy-form}
ds^2 =dz^2+\g_{\m \n}(z,x) dx^\m dx^\n~,\qquad \g_{\m\n}(z,x) = p^2(z)h_{\m\n}(x)~,
\ee
for some function $p (z)$ and
$h_{\m\n} (\m,\n = 0\cdots, d-1)$ is the boundary metric.
The boundary is supposed to be at some location $z = a$ in this
coordinate system.

\subsection{Scalar case} \label{b2b-scalar}

First let us consider the case of a real scalar field
with the action
\be
I(\phi)=-\frac{1}{2} \int d^{d+1}y \sqrt{g} (g^{MN}
\partial_M\vphi \partial_N \vphi+m^2 \vphi^2),
\ee
where $g$ is the absolute value
of the determinant of the matrix $g_{MN}$.
Performing an integration by parts leads to
\be\label{act1}
I(\phi)=-\frac{1}{2} \int_{\del M} d^{d}x\sqrt{\g}\vphi n^M\partial_M\vphi
+\frac{1}{2} \int d^{d+1}y\sqrt{g} \vphi(\Box-m^2)\vphi,
\ee
where $n^M:=\partial x^M/\partial z$
is the normal vector to the boundary surface. For our metric
\eq{bdy-form}, the only nonvanishing component is $n^z =1$.
The first term of \eq{act1} is a  boundary term and the second term gives
the equations of motion
\be\label{kg1}
(\Box-m^2) \vphi=0~,
\quad \Box :=\frac{1}{\sqrt{g}}\partial_M(\sqrt{g} g^{MN}\partial_N \; ),
\ee
where  $\Box$ is the $(d+1)$-dimensional d'Alembertian operator of $\cM$.

When evaluated on shell, the bulk term in \eq{act1} vanishes and
only  the boundary term contributes. It is clear that the boundary contribution
depends on the boundary behavior of the solution of the equation of
motion.  Without loss of generality, consider a solution of  $\vphi$ with the
following leading asymptotic behavior near the boundary,
\be
\vphi \sim f(z) \vphi_0(x) , \quad z \sim a,
\ee
for some function $f$ of $z$. It is then convenient to introduce the
bulk to boundary propagator $K$ defined by the following differential
equation
\be \label{K-def}
\Box K(z,x,x')=0~,
\ee
and has the boundary behavior
\be \label{K-def2}
 \lim_{z\rightarrow a} K(z,x,x')\sim f(z) \frac{\delta^{(d)}(x-x')}{\sqrt{h}}~,
\ee
where $h$ is the absolute value of the determinant of the boundary
metric $h_{\m\n}$
and $\sim$ denotes the leading order contributing term in the sense of
distribution.
For convenience, we have chosen to include a volume factor
$\sqrt{h}$ in our definition of $K$, making it scalar.
The introduction of the bulk to boundary propagator
allows us to write  $\vphi(z,x)$ as
\be\label{phi1}
\vphi(z,x)=\int d^d x'\sqrt{h(x')} K(z,x,x')\vphi_0(x')~,
\ee
in agreement with the definition \eq{K-def2}. If we substitute \eq{phi1} into \eq{act1}, we get
\be
I(\phi)=-\frac{1}{2} \int d^{d}x\sqrt{h(x)} d^{d}x' \sqrt{h(x')} \;
\vphi_0(x) \cG(x,x') \vphi_0(x') ~,
\ee
with
$\cG(x,x')$ defined as
\be\label{cg1}
\cG(x,x'):=\lim_{z\rightarrow a}
\prt{f(z) p^d(z) \partial_z  K(z,x,x')} ~.
\ee
According to the prescription of AdS/CFT correspondence \cite{witten,GKP},
the 2-point function of the dual field theory is given by $\cG (x,x')$.
The equation \eq{cg1} expresses it in terms of the normal derivative
of the bulk-to-boundary propagator $K$ at the boundary.

The key goal now is to obtain the bulk-to-boundary propagator. In the
original works \cite{witten,GKP}, this is obtained by solving the
differential equation \eq{K-def} directly. There is however a more
effective way.
Let us introduce the Green function for the bulk
\be \label{green}
(-\Box+m^2)G(z,x;z',x')=\frac{1}{\sqrt{g}}\delta^{(d)} (x-x')\delta(z-z')~.
\ee
Using the Green's identity, one can easily obtain
the solution of the scalar Klein-Gordon equation \eq{kg1}
in terms of the Green function as
\be
\vphi(z,x)=\int d^{d} x'\vphi_0(x')  \sqrt{\g(z',x')} \Big(
G \del_{z'} f(z') -f(z')  \partial_{z'} G
\Big)_{z'\rightarrow a}~.
\ee
Comparing with \eq{phi1}, the bulk-to-boundary propagator $K$ can be
written in  terms of the Green function as
\be\label{ktog}
K(z,x,x')= \lim_{z' \rightarrow a}  p^d(z')
\Big(
G(z,x;z',x') \del_{z'} f(z') -f(z') \partial_{z'} G(z,x;z',x')
\Big)
\ee
and subsequently the two point function $\cG(x,x')$ can be obtained
in terms of $K$ using \eq{cg1}.
This formula displays clearly how the bulk physics, as encoded in the
bulk Green function, is translated to the physics on the
holographic field theory
through the boundary data: the asymptotic behavior $f$ of
the field and of the
metric volume factor $p^d(z)$  near the boundary. 
The relation of the propagator with the Green
function and its derivative, turns out to be 
always 
scaled in such a way 
that the limit at the boundary is finite.
Higher point functions can be obtained by the Witten diagrams \cite{witten}.

\subsection{Spin 1/2 case} \label{b2b-fermion}

The action for a massive free spin 1/2 fermion
on our $(d+1)$-dimensional space $\cM$ reads
\be\lab{actionf}
I_0=\int d^{d+1} x \sqrt{g} \psib\prt{\Dss-\m}\psi,
\ee
where $g_{MN} $ is the metric on $\cM$. Without loss of generality, we
assume $\m>0$. The Dirac equation of motion reads
\be\lab{dir1}
\prt{\Dss-\m}\psi=0~, \quad \psib (- \overleftarrow{\Dss} -\m) =0.
\ee
The action \eq{actionf} vanishes on-shell. For the application of
AdS/CFT, one needs to supplement it with the boundary action \cite{sfetsos}
\be
\frac{1}{2}\int_{\del \cM} d^d x\sqrt{\g}  \psib \psi ,
\ee
where $\del M$ is the boundary of $\cM$ and the metric takes the form
\eq{bdy-form} near the boundary. In practical calculation,
as various quantities in $I_b$ are divergent near the boundary,
one needs to consider it as the  limit
\be \lab{i1}
I_b := \lim_{z \to a} \frac{1}{2}\int_{\del \cM} d^d x\sqrt{\g}  \psib \psi .
\ee
The necessity of the boundary term was justified in  \cite{henneaux}
which demonstrated that only then the variational principle for the
fermionic action  is well defined:
it ensures that by decomposing the spinor $\psi$ in terms of the
eigenvalues
of say, the Gamma matrix $\G^z$, the on-shell action is not a function
of both components, since the regularity of the solution on $\cM$
restrict that only half the components of the spinor $\psi$ can be
prescribed on the boundary.

Now let us consider a solution $\psi$ with the leading asymptotic behavior near
the boundary
\be
\psi \sim f(z) \psi_0(x), \quad z\sim a,
\ee
where $f(z)$ is a function and  $\psi_0$ is a spinor living on the boundary.
It is easy to see that for positive $\m$, the non-normalizable mode
is obtain from $\psi_0$ of negative chirality:
\be
\G^z \psi_0 =- \psi_0.
\ee
The fermionic bulk to boundary propagator $S(z,x,x')$  from a point
$(z,x^\m)$ in the interior to a point $x'{}^\m$ on the boundary
is defined by the
differential equation
\be \label{S-def1}
(\Dss -\m) S =0
\ee
and the boundary behavior 
\be \label{S-def2}
\lim_{z \to a} S(z,x,x') \sim f(z) \frac{\d^{(d)}(x-x')}{\sqrt{h}} \id.
\ee
It allows us to write the on-shell configuration of $\psi$ as
\be\lab{sola1}
\psi(z,x)=\int d^d x' \sqrt{h(x')}S\prt{z,x,x'} \psi_0\prt{x'}~,
\ee
in agreement with the definition \eq{S-def2}.
Substituting \eq{sola1} into $I_b$, we obtain
\be\lab{i2}
I_b=\int d^d x'\sqrt{h(x')} d^d x''\sqrt{h(x'')}\;
\psib_0(x')\cG(x',x'') \psi_0(x'')~,
\ee
where
\be \label{G2-fermion}
\cG(x',x'') := \lim_{z\to a} \frac{1}{2}
\int d^d x \sqrt{\g(z,x)} S^\dag(z,x,x') S(z,x,x'')~.
\ee
As $S$ behaves like a delta function near the boundary, the integral
picks up its contribution from the two regions: $x \sim x'$ and $ x\sim
x''$ and we obtain
\be \label{cg2}
\cG(x',x'') = \lim_{z\to a} \frac{1}{2} f(z) p^d(z) \prt{  S^\dag(z,x'',x')
+   S(z,x',x'') }.
\ee
This formula is analogous to \eq{cg1} and gives the fermionic
two point function
in terms of the fermionic bulk to boundary propagator $S$. To find
$S$, one can try to solve for it directly from the defining equations
\eq{S-def1} and \eq{S-def2}.  This has been carried out in
\cite{sfetsos,mueck,henneaux} for the original AdS/CFT correspondence
with Minkowski CFT living on the boundary.
However this is not necessary as we only
need $S$ near the boundary in \eq{cg2}.
In the next section, we will show that  $S$ in \eq{cg2}
can be obtained in terms of the
scalar bulk to boundary propagator $K$:
\be \label{cg3}
S = \Dss K.
\ee

As noted in \cite{pope}, apart from having Minkowski spacetime M${}_4$ and de
Sitter spacetime dS${}_4$ as boundaries, it is also possible to consider another
coordinate patch of the AdS${}_{5}$ and have AdS${}_4$ as
boundary. In this case boundary dual theory is a superconformal
Yang-Mills theory living on AdS${}_4$. The general
results obtained  here can be applied to all these cases.

\section{Scalar 2-Point Function in dS Dual Field Theory}

In the above we have shown how the bulk to boundary propagator and the
two point correlators in the boundary field theory can be derived from
a knowledge of the bulk Green function.
In this section, we consider different boundaries of the AdS
space and will use these formulas to show how the same AdS bulk physics
could manifest itself differently on the boundary field theories.

In the standard AdS/CFT correspondence which employ the Poincare
coordinate
patch \eq{adsp}  of the AdS space, one can put the metric in the
form \eq{bdy-form}  by writing $ r = \exp(-z/L)$. It is
\be
\sqrt{-\g } = \frac{1}{r^d}=  e^{dz/L}.
\ee
Near the boundary $r' \to 0$ or $z'\to \infty$,
from \eq{geod-poin} we have
\be
\xi \approx \frac{2 r r'}{r^2 + |x-x'|^{2} } \to 0.
\ee
To compute the bulk to boundary propagator $K$ from \eq{ktog},
we need to know the Green function near the
boundary. For AdS${}_{d+1}$ in the Poincare coordinates, we have, see
for example \cite{r2},
\be\lab{greenads}
L^{d-2}G(X,X')=\frac{C_{\D}}{2\n}\prt{\frac{\xi}{2}}^{\D}
F\prt{\frac{\D}{2},\frac{\D+1}{2},\n+1;\xi^2}~,
\ee
where
\be
C_{\D}:=\frac{\G(\D)}{\pi^{\frac{d}{2}}\G\prt{\D-\frac{d}{2}}}
\ee
and $\D$ can be either $\D_+$ or $\D_-$
\be
\D_\pm := \frac{d}{2}\pm \n~,\quad \n:=\sqrt{\frac{d^2}{4}+m^2 L^2}~ .
\ee
Using the properties
\be
\lim_{x\to 0} F(a,b,c;x) =1~, \quad
\frac{d}{dx}F(a,b,c;x) =\frac{ab}{c} F(a+1,b+1,c+1;x)~,
\ee
of the hypergeometric functions, we get
\bea\nn
\pa_{z'} \prt{\xi^{\D} F}= -\frac{\D}{L} \xi^{\D}  +\cO\prt{\xi^{\D+1}}
\eea
near the boundary.
Now it is the non-normalizable mode  with the asymptotic
behavior $f = r^{\D_-}$ that defines a field at the boundary.
It is easy to see that in order to
get the desired boundary behavior \eq{K-def2}  for $K$, we need to
adopt the root $\D =\D_+$ in the Green function \eq{greenads} and
we obtain the bulk to boundary propagator (up to a constant)
\be
K = \Big(\frac{r}{r^2+ |x-x'|^{2}  } \Big)^\D.
\ee
It follows immediately from \eq{cg1} that
\be
\cG(x,x')=  \frac{1}{|x-x'|^{2\D}},
\ee
which is the expected form of the two point function for operator of
dimension $\D$ in  CFT living on Minkowski spacetime.

Next we consider the  dS slicing \eq{ds1} of AdS.
To compute the 2-point function in the corresponding boundary
conformal
field
theory, we use the fact that the Green function
is a scalar and therefore invariant under
coordinate transformations. The AdS${}_{d+1}$ metric $g$ we consider is
\be
ds^2=dz^2+\sinh^2 (Hz)\, ds^2_{\rm dS}~,
\ee
and has a dS boundary with metric $h$, where
$\sqrt{g}=\sinh^d(Hz)\sqrt{h}$.
Near the boundary $z' \rightarrow \infty$, \eq{geod-planar} gives
\be\lab{sdef}
\xi=\frac{e^{-H\prt{z+z'}}}{H^2 \r^2/8} \rightarrow 0 ,
\ee
where
\be
\r^2 = \s^2 + e^{-2Hz}\cdot \frac{1-H^2\s^2/4}{H^2/4}
\ee
and $\s^2$ is given by \eq{sigma1}.
Working similarly as above, we need the asymptotics of the
non-normalizable mode near
the boundary $z \to \infty$ in the dS slicing. In our coordinates,
the  Klein-Gordon operator $\Box$ is given by
\be
\Box=\pa_z^2+ \frac{d H}{\tanh H z}\pa_z+\frac{1}{\sinh^2 H
  z}\Box_{\rm dS}, \quad \Box_{\rm dS} =
\frac{1}{\sqrt{-h}}\pa_\m\prt{\sqrt{-h} h^{\m\n}\pa_\n}.
\ee
Consider a solution of the form $\vphi\prt{z,x}=f(z) \vphi_0 (x)$,
near the boundary we obtain the second order ordinary differential equation
\be
f''+ d Hf'-m^2 f=0~,
\ee
with the non-normalizable solution
\be\lab{solkg}
f=e^{-\D_- H z}~,\qquad \D_\pm:=
\frac{d}{2}\pm\sqrt{\frac{d^2}{4}+\frac{m^2}{H^2}}~.
\ee
The bulk to boundary propagator is given by (up to a constant),
\be
K = \Big(\frac{e^{-Hz}}{\r^2}\Big)^\D, \quad
\mbox{with $\D = \D_+$}
\ee
and we obtain from \eq{cg1} the two point function
\be \label{res1}
\cG (x,x') =  \frac{1}{\s(x,x')^{2\D}}.
\ee
This is the expected form of the two point function for operators of
dimension $\D$ of a  conformal field theory in  dS spacetime.

\section{Fermion 2-Point Function in dS Dual Field Theory}

\subsection{Flat space conformal field theory}

Let us first consider the canonical case \eq{adsp} of AdS in Poincare
coordinates,  with Minkowski spacetime located at the boundary at
$r=0$. The vielbein reads
\be
e^A_M=\frac{1}{r}\d^A_M~,
\ee
with the curved indices $M=(r,\m)$ and flat indices $A=(r,a)$
having range $a,\m=0,\ldots,d-1 $. For convenience,
we will assume $L=1$ in this
subsection. $L$ can be restored easily by dimensional analysis.
The Dirac operator is given by 
\be
\Dss= e^M_A \G^A(\del_M + \frac{1}{2}\o_M{}^{BC} \S_{BC}) =
r \G^r \del_r+r\G^\m \del_\m-\frac{d}{2}\G^r~,
\ee
where the matrices $\G^A = (\G^r,\G^\m)$ are the flat space gamma matrices:
\be
\{\G^A,\G^B\}=2\eta^{AB}~, \quad \eta_{AB} = {\rm diag} (1, \eta_{d}).
\ee
The Dirac equation \eq{dir1} in AdS${}_{d+1}$ reads
\be\lab{dir2}
\prt{\G^r\prt{r \del_r-\frac{d}{2}} +r \G^\m\del_\m-\m}\psi=0 ~.
\ee
The asymptotic behavior of the on-shell mode is govern by the
behavior of the Dirac operator near the boundary, we obtain
\be
\psi \sim r^{\tD_-} \psi_0(x), \quad r\to 0,
\ee
where
\be
{\tD}_\pm := \frac{d}{2}\mp \m \G^r.
\ee
In order for this mode to be non-normalizable, it is necessary to take the
boundary spinor $\psi_0$ to be of
negative chirality
\be
\G^r \psi_0 = - \psi_0
\ee   and $\m > d/2$
such that $\tD_- <0$.

To construct $S$,  it is useful to note that the Dirac operator
$\Dss$ satisfies the following relation
\be \label{Dss2-flat}
\Dss^2=\prt{\Box+\frac{d^2}{4}} \id-r\G^\m\G^r\del_\m~,
\ee
where
\be
\Box=r^2\del_r^2+r(1-d)\del_r+r^2\del_\m^2~
\ee
is the  d'Alembertian  on AdS${}_d$ space in the metric \eq{adsp}.
\eq{Dss2-flat} can be written in the more convenient form
\be
\Dss^2+\prt{\Dss-\m}\G^r=\prt{\Box+\frac{d^2}{4}}+\prt{
r\del_r-\frac{d}{2}-\m \G^r}~.
\ee
We propose  to construct the fermion bulk
to boundary propagator  $S$
as follow:
\be \label{pr1}
S=\prt{\Dss+\m +\G^r}K+\d~,
\ee
where
\be
K = \Big(\frac{r}{\r^2} \Big)^{\D_+}
\quad \mbox{with} \quad
\r^2 := r^2 + (x-x')^2 ~,
\ee
is the bulk
to boundary propagator for an auxiliary scalar field of mass $m$;
\be \label{pr2}
\d:=- \prt{\Dss-\m}^{-1}\prt{r\del_r-\tilde{\D}_-}K~.
\ee
We remark that
our notation for $\tD_\pm$ takes into account that our boundary spinor $\psi_0$
lives in the $\G^r=-1$ sector.
It is easy to see that in order for $S$ to satisfy the equation of
motion \eq{S-def1}, $m$ has to be given by
\be \label{pr4}
m^2= \m^2 -\frac{d^2}{4}.
\ee
This relation is also 
precisely 
what is needed to guarantee that $S$ satisfies  the
desired boundary condition. In fact, since $K$ has the
boundary behavior 
\be
\lim_{r\to 0} K \sim r^{\D_-} \d^{(d)}(x-x').
\ee
As a result, $\d$ vanishes at the boundary
since $\D_- = \tilde{\D}_-$ (due to \eq{pr4} and
$\G^r =-1$ when acting on $\psi_0$). Now, it is easy to check that
\be \label{DssK}
\Dss K = \m K - 2 \tilde{\D}_- \frac{r}{\r} U K~,
\ee
where
\be
U := \frac{r \G^r + (x-x')_\m\G^\m}{\r}~.
\ee
Note that $U^2 = \id$.
As one approaches the boundary, $K$ imposes $x=x'$ implying
\be
\frac{r}{\r} \to 1, \quad U \to \G^r,
\ee
and so $S$
satisfies the same (up to proportional constant) boundary condition as $K$:
\be
\lim_{r \to 0} S \sim r^{\tD_-}  \d^{(d)}(x-x').
\ee

In principle one can use \eq{pr1} to compute the full expression of
$S$. However for our purpose of computing the two point function using
\eq{cg2}, this is not necessary as  we only need to know $S$
near the boundary.
Note  that the $\mu$ and $\G^r$ term of \eq{pr1}
do not contribute to the boundary action \eq{i2} since
\be\lab{cons}
\psib_0 \psi_0 =0~,\qquad \psib_0 \G^r \psi_0= 0,
\ee
where we have used the fact that $\psi_0$ and $\psib_0$ has opposite
chirality. Also, since $\d =0$ at the boundary.
As a result, the form of $S$ to be used in \eq{cg2} is given by
\be
S= \Dss K.
\ee
Since only the second term in \eq{DssK} contributes, so
we obtain the fermionic two point function
in  the conformal field theory living on the boundary flat space
\be\lab{fermiong11}
\cG(x,x')= \frac{(x-x')_\m \Gamma^\m}{|x-x'|}
\frac{1}{|x-x'|^{2 \tilde{\D}_+}}~.
\ee
The result \eq{fermiong11} was first obtained by \cite{sfetsos}
using the AdS/CFT correspondence.

\subsection{dS conformal field theory}

Next let us consider the AdS${}_{d+1}$ space with a $dS$ boundary with
the metric
\be\lab{metric1}
ds^2=dz^2 +\sinh^2 H z\prt{-dt^2+e^{-2 H t}dx_i^2}.
\ee
The vielbein $e^M_A$ is
\be
e_{z}^M=\d^M_z~,\quad e_t^M=\frac{1}{\sinh Hz} \d_t^M ~,\quad
e^M_i=\frac{e^{H t}}{\sinh H z} \d_i^M~.
\ee
The non-zero spin connection elements $\o_M{}^{AB}$ are
\be
\omega_M{}^{t~ z}=H \cosh H z \delta_M^t~,\qquad
\omega_M{}^{i~ z}=H \cosh H z e^{-H t} \delta_M^i~,\qquad
\omega_M{}^{i~ t}=H e^{-H t} \delta_M^i~
\ee
and the Dirac operator reads
\be
\Dss=\G^z \partial_z +\frac{1}{\sinh H z}\prt{\G^t\del_t+e^{Ht} \G^i
  \del_i}
- \frac{H\prt{d-1}}{2 \sinh H z} \G^t+\frac{d H}{2 \tanh H z} \G^z.
\ee
We note in passing that
\be
\Dss=\G^z \partial_z +\frac{1}{\sinh H z} \Dss_{\rm dS} +
\frac{d H}{2 \tanh H z} \G^z.
\ee
The asymptotic behavior of the solution to the Dirac equation can be
easily worked out. It is
\be
\psi \sim e^{-\tD_- H z} \psi_0(x), \quad z \to \infty~,
\ee
where here
\be
{\tD}_\pm := \frac{d}{2}\pm \m \G^z~.
\ee
 In order for this mode to be non-normalizable, we need
\be
\G^z \psi_0 = \psi_0
\ee
and $\m>d/2$.

The general solution of the Dirac equation can be written in the form
\eq{sola1} with the boundary spinor $\psi_0$ satisfying $\G^z \psi_0=\psi_0$.
As before we note the useful relation:
\bea \label{DD-ds}
&&\Dss^2-(\Dss -\mu) H \coth H z ~\G^z
= \prt{\Box+\frac{d^2 H^2}{4 \tanh^2 H z}
-\frac{\prt{d-3}\prt{d+1} H^2}{4 \sinh^2 H z }} \nn \\
&&\qquad\qquad\qquad\qquad
-H \coth H z \prt{\del_z - \mu \G^z +
\frac{d H}{2 \tanh H z}}+\frac{H e^{Ht}}{\sinh^2 H z}\G^t \G^i \del_i~,
\eea
where
\be
\Box=\del_z^2 -\frac{1}{\sinh^2 H z}\del_t^2+
\frac{e^{2 H t}}{ \sinh^2 H z}\del_i^2+\frac{d H}{\tanh H z}\del_z
+\frac{(d-1)H}{\sinh^2 H z}\del_t ~
\ee
is the  d'Alembertian on AdS${}$ space with the metric \eq{metric1}.
This allows us to construct the fermion bulk to
boundary propagator $S$ as
\be \label{S-ds}
S =\prt{\Dss+\m-H \coth H z~ \Gamma^z}K +\d~,
\ee
where
\be
K = \prt{\frac{e^{-Hz}}{\rho^2}}^{\D_+}
\ee
is the bulk to boundary propagator
for an auxiliary scalar field of mass $m$;
\be
\d:=-\prt{\Dss-\m}^{-1}\prt{q(z)-H \coth Hz
\prt{\del_z + H \tilde{\Delta}_-} +\frac{H e^{Ht}}{\sinh^2 Hz}\G^t
\G^i \del_i  }K~,
\ee
and
\be
q(z):=\frac{d H^2}{2}\frac{\tanh H z-1}{\tanh^2 H z}-
\frac{\prt{2d+3} H^2}{4}\frac{1}{\sinh^2 H z }~.
\ee
It is easy to see that one needs again \eq{pr4} in order for $S$ to
satisfy its defining differential equation. One also easily see that $\d$
vanishes at  the boundary and $S$ satisfies the desired boundary
condition.

To calculate the boundary two point function using \eq{cg2}, we
only need to know $S$ near the boundary, which is
$S = \prt{\Dss  + \m - H \coth H z~ \Gamma^z} K $.  As terms
that are proportional to the unit matrix or $\G^z$ do not
contribute to \eq{i2}, we can drop them in $S$ and we obtain again the relevant
expression \eq{cg3} for $S$. In terms of conformal coordinates,
it is
\be\lab{sk}
S =\frac{1}{2}
 e^{- H z}\, x_0 \G^\m \del_\m K -\frac{H(d-1)}{2}
\G^0 e^{- H z} K \quad\mbox{for large $z$}.
 \ee
Now the first term of \eq{sk} is equal to 
$-\D_+ b K D$,
where $b$ is the function
$ b:= e^{-H z} / \r$
and  $D$ is the matrix
$ D := x^0 x^\m \del_\m \s^2/(2 \s) $.
Near the boundary, $K$ imposes $\s =0$ and so $b=2/H$
and
\be
D = \frac{(x-x')_\m \G^\m}{|x-x'|}.
\ee
It is then
clear that the second term in \eq{sk} is sub-leading compared to the
first term. As a result,  the boundary two-point function is given by
(up to a constant)
\be
\cG\prt{x,x'}=
\frac{D}{\s^{2\tD_+}}~.
\ee
This agrees with the result
\eq{OOD} 
for a  conformal field theory in dS spacetime.

\section{Discussion}

In this paper, we have proposed  an AdS/CFT
correspondence of the Type IIB string theory on $AdS_5 \times S^5$ in
terms of a 
superconformal field theory on the dS${}_4$ boundary. As a first step, we have
provided evidence of it
by showing that  the boundary correlators 
of the dS conformal field theory
can be
reproduced from the AdS bulk dynamics.
It would be interesting to compute the higher point functions and see
if there is any nonrenormalization theorem for chiral operators as in
\cite{seiberg}.
It would also be interesting to look
at minimal surfaces 
with different boundary conditions 
and compute the interacting potential between quarks
anti-quarks and 
study properties of the entanglement entropy in a quantum field
theory in a time dependent background.

According to the holographic principle
\cite{hooft,susskind} (see also \cite{bousso} for a review),
the full description of quantum gravity in a
region requires only a quantum field theory  living at the
boundary.
Gauge/gravity correspondence is a nice illustration of the
holographic principle. In the original works of \cite{witten, GKP}, a
boundary flat $R^4$ was created when the Poincare patch of the AdS
spacetime  
was considered. 
In this paper we considered a
different coordinate patch with a different boundary, and showed
that the same bulk dynamics (e.g. the same
IIB supergravity equations in the classical supergravity limit),
with different boundary conditions
would results in different
holograms.
This is of course 
in consistent with  the statement
of the holographic principle. Nevertheless this is an aspect of the
holographic principle that
has not been emphasized much so far in the literature.
In the full non-nonperturbative formulation of string theory on
$AdS_5 \times S^5$, all the
different boundaries and boundary conditions should be contained in the
moduli space of the theory itself. This implies that the different dual quantum
field theories may also be considered  to be contained in some bigger
theory of quantum field theories.

On the $AdS_5 \times S^5$  side, the existence of the Lax pair and an
infinite set of
classically conserved nonlocal charges are properties of the
Green-Schwarz
string sigma model
\cite{polchinski}. Note that these properties were established
using the global AdS space without referencing to the Poincare
coordinate system. As such, the construction \cite{polchinski}
of charges also apply in
our case. One may speculate that the $\cN=4$
superconformal Yang-Mills on dS${}_4$ may also be integrable in some of
its sectors. This is an
interesting aspect that deserves further studies.


\vskip7mm
\section*{Acknowledgements}

It is our pleasure to thank Jean-Pierre Derendinger for many
useful discussion and
collaboration in the early stage of this work.
CSC also thank Koji Hashimoto and Piljin Yi for helpful
discussion and suggestion.
This work is
supported in part by  the National Center of Theoretical Science
(NCTS) and the grant
104-2112-M-007-001 -MY3 of the Ministry of Science and
Technology of Taiwan.


\end{document}